\documentclass[preprint,tightenlines,aps,preprintnumbers,showpacs,nofootinbib,floatfix]{revtex4}
\usepackage{amsmath,amssymb,bm,epsfig}

\newcommand\pd{\partial}
\renewcommand\Re{{\rm Re\,}}
\renewcommand\Im{{\rm Im\,}}
\newcommand\muB{{\mu}}
\newcommand\muc{{\mu_c}}
\newcommand\muR{{\mu_R}}

\newcommand\mq{{m}}

\begin{document}
\title{QCD Critical Point and Complex Chemical Potential Singularities}
\author{M.~A.~Stephanov}
\affiliation{Department of Physics, University of Illinois, 
Chicago, Illinois 60607-7059,USA}

\begin{abstract}
The thermodynamic singularities of QCD in the plane of {\em complex}
baryo-chemical potential $\muB$ are studied. Predictions are made
using scaling and universality arguments in the vicinity of the
massless quark limit. The results are illustrated by a calculation of
complex $\muB$ singularities in a random matrix model at finite
temperature.  Implications for lattice QCD simulations aimed at
locating the QCD critical point are discussed.
\end{abstract}
\date{March 2006}
\pacs{12.38.Gc 64.60.Fr 12.38.Mh}

\maketitle

\section{Introduction}

The search for the QCD critical point has attracted considerable
theoretical and experimental attention recently. The existence of such a
point -- an ending point of the first order chiral transition in QCD --
was suggested a long time ago \cite{Asakawa,Barducci},
and the properties were studied using universality arguments and
model calculations more recently
\cite{Berges,Halasz-pdqcd} (see Ref.~\cite{cp-review} for review).
The experimental search for the critical point 
using heavy ion collisions has been
proposed in \cite{signatures}. It is apparent that theoretical
knowledge of the location of the critical point on the phase diagram
is important for the success of the experimental search.

First principle lattice QCD calculations aimed at determining the
location of the critical point on the $T,\,\muB$ 
(temperature, baryon-chemical potential) diagram have been
attempted recently using several different techniques
\cite{Fodor,Allton,Gavai,deForcrand} (see Ref.\cite{Philipsen-review}
for review).  The major obstacle for direct
Monte Carlo simulation is the well-known lack of positivity of the
measure of the path integral defining QCD partition function at
nonzero baryo-chemical potential $\muB$ --- the sign
problem. One of the methods to deal with this problem is to Taylor
expand the QCD pressure in powers of $\muB$ around $\muB=0$, i.e., around
the point at which direct Monte Carlo simulations are not hindered by
the sign problem \cite{Allton,Gavai}.

The success of such an approach in determining the location
$(T_E,\,\mu_E)$ of the critical ending point crucially depends on the
convergence radius of the Taylor expansion around $\muB=0$
\cite{Allton,Gavai,Lombardo}. The convergence radius, in turn, is a
function of the position of the singularities in the complex $\muB$
plane. Little is known about the location of these singularities to
date. The purpose of this paper is to expand our knowledge of the
location of these complex plane singularities.

We shall be able to determine the position of the singularities 
in the regime where the quark masses $\mq$ are sufficiently small.
To summarize, the strongest rigorous consequence of
our analysis is that the convergence radius, $\muR$,
achieves its minimum value at a temperature slightly
above $T_c$ (by ${\cal O}(m^{1/(\beta\delta)})$). This value scales as
\begin{equation}\label{mur}
  \min_T\, \mu_R(T) \sim m^{1/(2\beta\delta)},
\end{equation}
and vanishes in the chiral limit ($m\to0$).
The value $1/(\beta\delta)\approx 0.54$ is determined by the
critical exponents of the O(4) universality class in 3
dimensions. The singularity which determines
the radius in \eqref{mur} lies in the complex plane and pinches the real axis
(together with its conjugate) at the critical point. The convergence
radius $\mu_R(T)$ has a certain nonanalyticity at $T=T_E$, which
we shall describe.

The study of thermodynamic singularities, or partition function
zeros, in the complex plane was pioneered by Yang and Lee 
\cite{YangLee}. Their analysis was extended by Fisher
from the complex magnetic field singularities to complex temperature
singularities \cite{Fisher}.  
The properties of these singularities following from
scaling and universality have been further studied by Fisher
\cite{Fisher-edge}, Itzykson,
Pearson and Zuber \cite{Itzykson} and others.

From the point of view of the lattice studies of the QCD phase
diagram, we would like to distinguish two separate issues. One is the
convergence of the series as the truncation order is increased.  This
is the issue which this study will impact.  The other is the
convergence of each term in the series to its thermodynamic limit as
the volume and/or the number of Monte Carlo configurations are increased.
The sign-problem is affecting this latter convergence and will not
be addressed here (see, e.g., \cite{Halasz-convergence,Splittorff,Ejiri}).

\section{Universal properties of the complex singularities}
\label{sec:universal}

Here we describe generic universal properties of the complex
thermodynamic singularities. The basic results following from scaling
and universality are not new, and can be found in \cite{Itzykson}. For
clarity and completeness, we rederive the needed facts here using slightly
different approach.
 Our purpose is to apply these results to QCD at
finite temperature and density.

\subsection{Complex $\mu$ singularities as Fisher zeros}

\begin{figure}
\centerline{
\epsfig{file=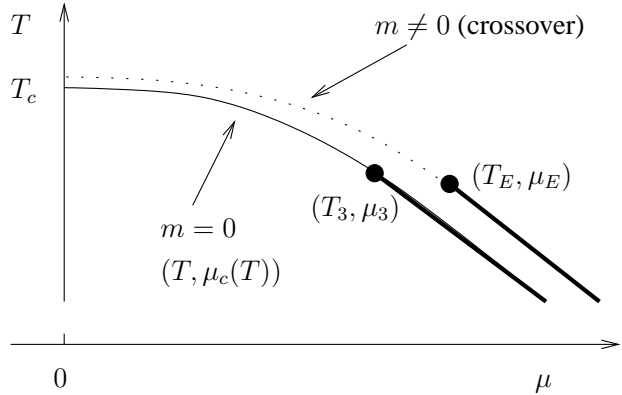,width=.5\textwidth}
}
  \caption[]{A sketch of the QCD phase diagram in the vicinity of the
critical line $\mu_c(T)$ for zero and nonzero quark mass.}
\label{fig:pd-sketch}
\end{figure}

We shall first consider QCD in the chiral limit -- the limit of two
lightest quark masses taken to zero. The phase diagram in the
$(T,\,\muB)$ plane is sketched in Figure~\ref{fig:pd-sketch}. The
low-temperature phase is separated from the high-temperature phase by
a phase transition. This is unavoidable, because the symmetry of the
ground state must change from SU(2)$_V\times$U(1)$_B$ to the full
symmetry of the action SU(2)$_V\times$SU(2)$_A\times$U(1)$_B$ as the
temperature is raised \cite{Kogut-book}.  This transition is of second
order for $\muB<\muB_3$, and of first order for $\muB>\muB_3$.

Let us now fix $T$ at a value in the interval $(T_3,T_c)$ and study
the behavior of singularities in the complex $\muB$ plane.  On the
real axis, as $\muB$ increases from zero, the transition occurs at a
value which we denote as $\muc(T)$ (see
Fig. \ref{fig:pd-sketch}). Therefore, at a fixed temperature, the
change of the chemical potential $\muB-\muc(T)$ is a relevant
perturbation. In the universality class of the O(3)$\to$O(4)
transitions, to which QCD chiral restoration transition belongs
\cite{Pisarski}, there is only one relevant variable -- the thermal
variable $t$ (we are discussing the symmetry limit -- the magnetic
field variable $h$ is absent). Therefore, we are led to consider the
universal behavior of the singularities in the complex {\em
temperature} plane, which correspond to Fisher zeros.  The scaling
parameter $t$ is to linear order
\begin{equation}\label{tmu}
  t\sim \muB^2-\muc^2(T).
\end{equation}
We use $\mu^2$ instead of $\mu$ to ensure that at $\mu=0$ the
thermal variable is proportional to $T-T_c$.

\subsection{Partition function zeros and electrostatic analogy}

As first exposed by Lee and Yang \cite{YangLee}, the thermodynamic
singularities in the complex plane are related to the zeros of the
partition function.  For a finite system the partition function $Z$,
by definition, is strictly positive for {\em real} values of the
parameters. However, $Z$ has zeros in the complex plane, whose number
grows linearly with the size of the system. In the thermodynamic limit the
zeros typically coalesce into cuts. A phase transition occurs where such
a cut pinches (second order) or crosses (first order) the real axis.

The considerations of Lee and Yang apply to the variable
$\lambda\equiv\exp(\muB/T)$ since the partition function of QCD
is a polynomial in this variable due to quantization of
the baryon charge.

It is convenient to use the electrostatic analogy. The
partition function -- a polynomial,
can be written in terms of its roots
\begin{equation}
  Z(\lambda)=\prod_k(\lambda-\lambda_k)\,.
\end{equation}
Therefore the free energy (or grand potential, if we are dealing with
grand canonical ensemble, as in QCD) is
\begin{equation}
  \Omega(\lambda) = -T\log Z = -T\sum_k\log(\lambda-\lambda_k).
\end{equation}
For complex $\lambda$, the real part $\Re\Omega$ can be interpreted as
an electrostatic potential created by charges located on the plane
$(\Re\lambda,\Im\lambda)$:
\begin{equation}
\Re  \Omega(\lambda) = -T\sum_i \log|\lambda-\lambda_k|.
\end{equation}
All charges have the same magnitude and sign (degenerate roots can be
treated as coincident charges). Now let us assume, as is known to be
true in most cases, and in the universality region of interest
in particular, that the zeros coalesce into 1-dimensional curves
in the thermodynamic limit.
Then, the electrostatic potential $\Re\Omega$ is continuous across
such a curve, while the analog of the electric field
\begin{equation}\label{E}
  \bm E = -\bm\nabla (\Re\Omega)=-\left(
\frac{\pd\Re\Omega}{\pd \Re\lambda},\,
\frac{\pd\Re\Omega}{\pd \Im\lambda}
\right)
= (-\Re,\,\Im)\,\frac{d \Omega}{d \lambda}
\end{equation}
is discontinuous -- the normal component jumps by an amount proportional
to the linear charge density $\rho$ on the curve. This curve can
be viewed as the location of a cut on a Riemann sheet of the
analytic function $\Omega(\lambda)$.

\subsection{Stokes boundaries in the scaling region at $h=0$}

Consider now the critical region in the vicinity of a
critical point $\lambda_c$. In the
scaling regime the singular (i.e., non-analytic) part of the
potential $\Omega(t)$ is proportional to a power of
$t$:
\begin{equation}\label{Osing}
  \Omega_{\rm sing}(t) = 
\left\{  \begin{array}{ll}
A_+\, t^{2-\alpha},&\quad t>0;\\
A_-\, (-t)^{2-\alpha},&\quad t<0;
  \end{array}
\right.
\,\qquad t\equiv\frac{\lambda-\lambda_c}{\lambda_c}
\end{equation}
which defines the universal specific heat exponent $\alpha$ and the amplitudes
 $A_{\pm}$, whose ratio is also universal. 
Off the real axis $\Omega(t)$ must be an analytic
 function everywhere except for discontinuities across the cuts.
Such cuts (at least two, by symmetry
 $\Omega(t^*)=\Omega(t)$) must be present, because the function
\eqref{Osing} taken at $t>0$ does not match its analytic 
continuation from the $t<0$ axis along a path around $t=0$. 
The location of the cuts can be
 determined using electrostatic analogy, which requires $\Re\Omega$
to be continuous across the cut. Parameterizing
 $t=-s\,e^{i\varphi}$ using real parameter $s>0$ we find:
\begin{equation}
  A_+\cos[(2-\alpha)(\varphi-\pi)] = A_-\cos[(2-\alpha)\varphi].
\end{equation}
Therefore, the cuts are straight lines at an angle
with respect to the negative $t$ axis given by (cf. \cite{Itzykson})
\begin{equation}\label{phi}
  \tan[(2-\alpha)\varphi] = \frac{\cos(\pi\alpha)-A_-/A_+}{\sin(\pi\alpha)}\,,
\end{equation}
as shown in Figure~\ref{fig:scalingzeros}.
All quantities entering this formula are universal.

The cuts are termed Stokes boundaries in \cite{Itzykson} 
--- they carry conceptual resemblance
to the anti-Stokes lines in the WKB theory. Across these lines,
the function $\Omega$ switches from one of its Riemann sheets to
another. The density of the ``charges'' on the cut is proportional to
the discontinuity of the normal component of $\bm E$, 
and thus to $\Im ( e^{i\varphi}\, d\Omega/dt)$, which vanishes at the branching point as
\begin{equation}\label{rho}
  \rho\sim |t|^{1-\alpha}.
\end{equation}

\subsection{Stokes boundaries at $h\ne0$}
\label{sec:hne0}

\begin{figure}
  \epsfig{file=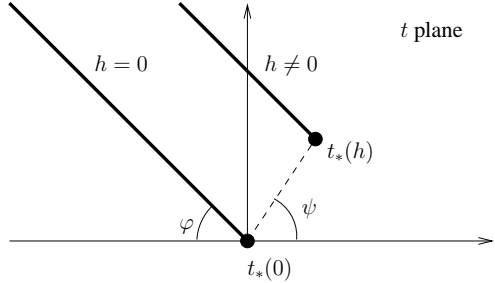,width=.4\textwidth}
  \caption[]{Universal behavior of Stokes boundaries
in the scaling region for zero and non-zero
  symmetry breaking parameter $h$. Only upper complex half-plane is
  shown. The trajectory of the branching
  point  $t_*(h)$ is indicated by a dashed line.}
\label{fig:scalingzeros}
\end{figure}

The magnetic field $h$ is another relevant variable near the O(4) critical
point. The magnetic field breaks the O(4) down
to O(3), and in QCD this role is played by the
quark mass $\mq$ (more precisely, the average of the $u$ and $d$ masses).
At $h\neq0$ the free energy is analytic function of $t$ at $t=0$.
In the scaling region the singular part of the free energy scales
as a power of $h$ if $t$ is also changed to keep 
the scaling variable $x\equiv t\,
h^{-1/(\beta\delta)}$ fixed.
Therefore the (two) branching points must be located at a point
away from the origin given by
\begin{equation}
  t_* = x_* h^{1/(\beta\delta)}
\end{equation}
and at $(t_*)^*$, where $x_*$ is a complex constant.
The phase of $t_*$ (the polar angle coordinate of the branching point) is
determined by the following argument. By scaling postulate,
the singular contribution to the free energy must be given by
\begin{equation}
  \Omega_{\rm sing} = h^{1+1/\delta} A\left(t\, h^{-1/(\beta\delta)}\right)
\end{equation}
where the function $A(x)$ is analytic at $x=0$. $A(x)$ has two
cuts which originate at points $x_*$ and $(x_*)^*$
and go off to infinity.
Consider $\Omega_{\rm sing}$ as a function of {\em complex} $h$ 
at fixed real~$t$.
By symmetry ($h\leftrightarrow -h$, $h\leftrightarrow h^*$),
the Stokes boundaries of this function lie
on the imaginary axis (more rigorously, this follows from the Lee-Yang
theorem and universality).
Thus the branching point, for $t>0$, at $h_*=(t/x_*)^{\beta\delta}$, 
is purely imaginary, and therefore $\beta\delta\arg x_*=\pi/2$.
Thus,
\begin{equation}\label{psi}
t_* = |x_*| e^{i\psi} h^{1/(\beta\delta)},
\qquad  
\psi=\frac\pi{2\beta\delta}\,.
\end{equation}


To summarize, at $h=0$, 
 the complex 
singularities in the scaling region of the thermal parameter $t$
form cuts (Stokes boundaries) which go along the rays at angle $\varphi$ with
the negative real axis given by \eqref{phi}.
With increasing symmetry
breaking parameter $h$ 
the branching point shifts away from $t=0$ by an amount proportional to
$h^{1/(\beta\delta)}$ 
along the direction at the angle
$\psi$ to the positive $t$ axis given by \eqref{psi}. This is
illustrated in Fig.~\ref{fig:scalingzeros}.

\section{Singularities of QCD in the complex $\muB$ plane}
\label{sec:qcd-universal}

\subsection{Chiral limit: $m=0$}

Let us begin with the chiral limit $\mq=0$. Consider the interval of $T\in
(T_3,T_c)$. In this interval increasing $\muB$ leads to the second order
transition at $\muB=\muc(T)$. As discussed in the previous section in
the vicinity of the transition we must identify $t$ with
$\muB^2-\muc^2(T)$ (up to an irrelevant constant factor) 
-- Eq.~\eqref{tmu}.
Thus, at a given temperature $T$, near $\muc(T)$ the location of 
the singularities in the complex $\muB$ plane
is determined by the universal arguments of the previous section.
Conformal transformation $\mu\to\mu^2$ does not affect the
angles $\phi$ and $\psi$ away from $\mu=0$.

In particular, at $m=0$, the (two) cuts should originate at the
branching point located at $\muc(T)$ on the real axis, and follow
the rays at angle $\varphi$  given by \eqref{phi} with respect to 
the negative real axis as shown in Figure \ref{fig:qcdzeros} (left).

Taking the values $\alpha\approx -0.25$ \cite{Pelissetto} 
and $A_+/A_-\approx1.6$ \cite{Toussaint} we estimate
the value of the angle as $\varphi\approx 77^\circ$. At the tricritical
point $\alpha=1/2$ and $A_+/A_-=0$ and thus $\varphi= 60^\circ$.

\subsection{Small quark mass: $\mq\ne0$}

At finite quark mass $\mq$ the second order line $\muB=\muc(T)$ is
replaced by an analytic crossover for all temperatures $T>T_E$.  The
critical ending point of the first order transition is located
at a temperature we denote $T_E$. For
$T<T_E$ the transition is of the first order.

At fixed $T>T_3$, and small mass $\mq$, in the vicinity of the crossover
 point the singularities are described by the universal arguments with
 \begin{equation}
   h\sim \mq\,.
 \end{equation}

\begin{figure}
\hspace{\stretch{.2}}
\epsfig{file=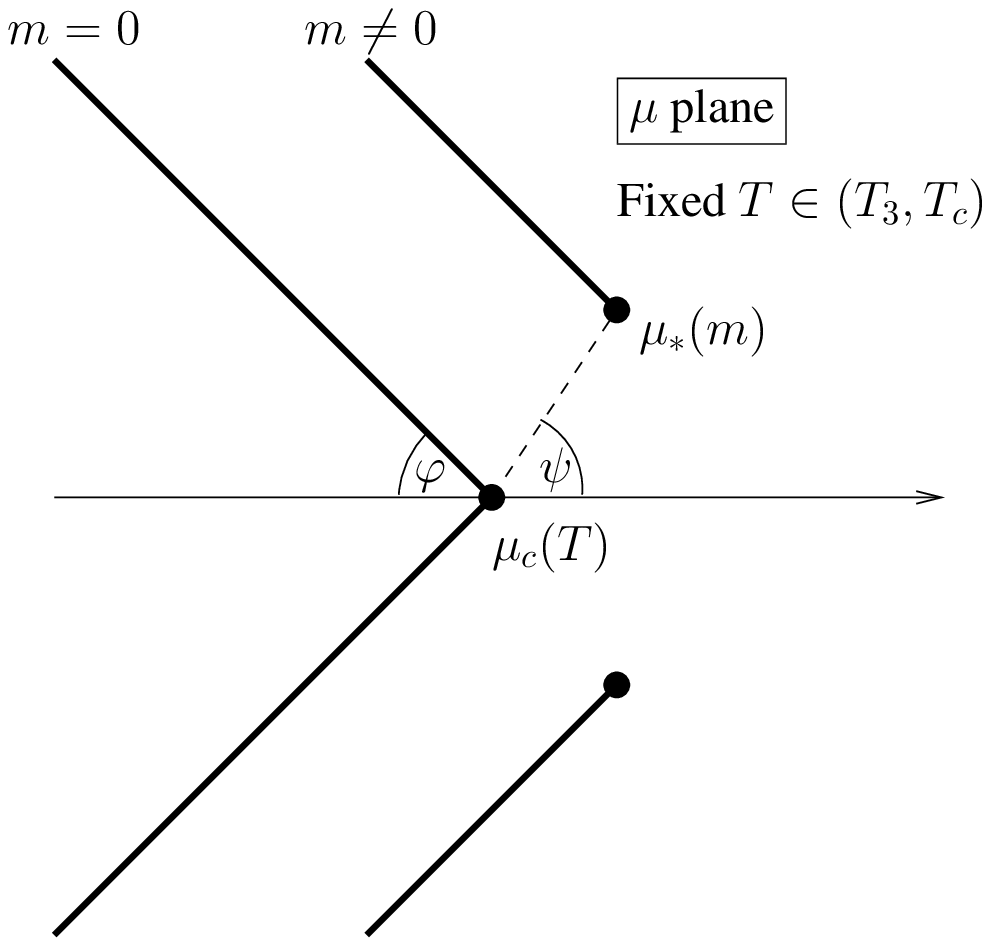,height=15em}
\hspace{\stretch{1}}
\epsfig{file=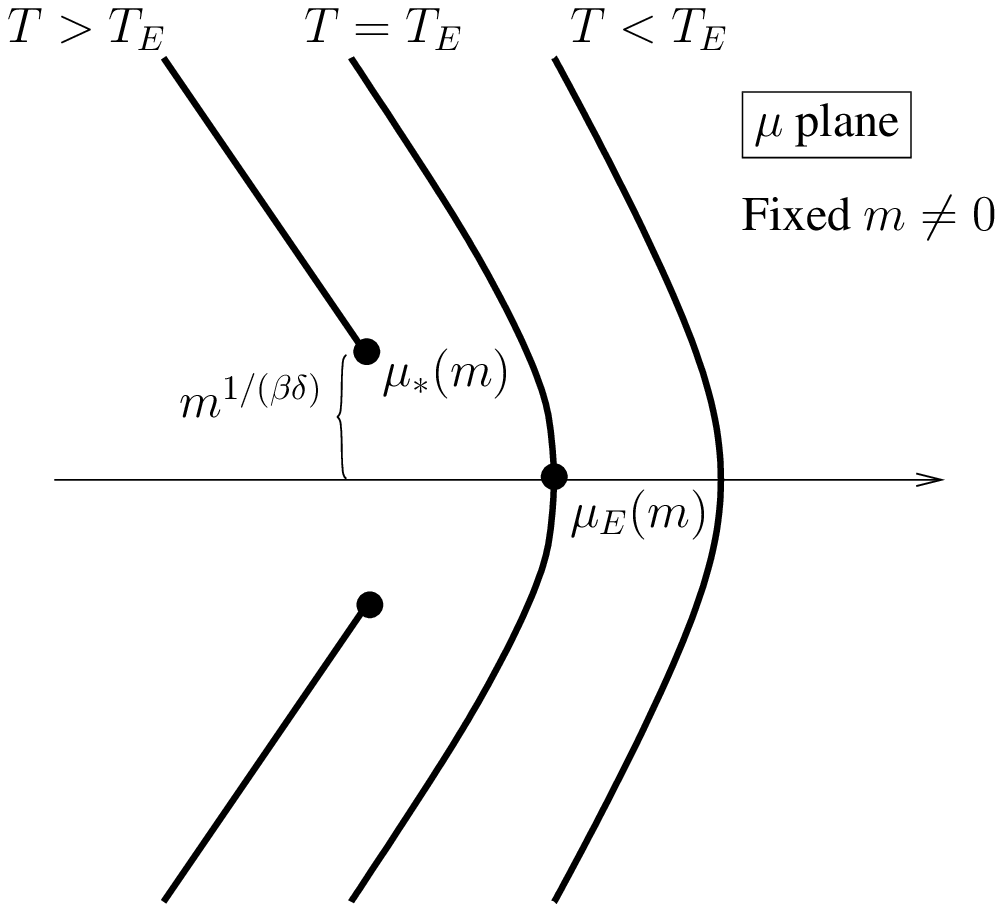,height=15em}
\hspace{\stretch{.2}}
   \caption[]{Stokes boundaries in QCD at fixed $T$ and two different values
     of $m$ (left) and at fixed small $m$ and three different values of
     $T$ (right) as
dictated by universality.}
\label{fig:qcdzeros}
 \end{figure}

For the purpose of the discussion we can define the crossover point as the
 value of the real part of the branching point: 
$\mu_{\rm crossover}\equiv\Re(\mu_*(m))$.
The branching point, $\mu_*(m)$ or the nearest
 singularity to the real axis, is shifted by the amount proportional
to $m^{1/(\beta\delta)}$, which is $m^{0.54}$, using the O(4) critical
exponents \cite{Pelissetto}, in the direction along the ray at angle 
\eqref{psi}
$\psi=\pi/(2\beta\delta)\approx 48^\circ$  towards the positive real
 axis  as shown in Figure~\ref{fig:qcdzeros} (left).

When $T$ is decreased towards $T_E$ this branching point $\mu_*(m)$
(and its conjugate) approach the real axis and pinch it when $T=T_E$. 
At this temperature, the cuts originate
from the branching point on the real axis $\mu_E$ (see Figure
\ref{fig:qcdzeros} (right)). The point
 $(T_E,\mu_E)$ is the QCD critical ending point. This ordinary critical
point is in the universality class of the 
Ising model~\cite{Berges,Halasz-pdqcd}. The initial
direction of the cuts near this point is perpendicular to the real
axis, i.e., $\varphi=90^\circ$.  This follows from the fact that the
perturbation $\muB-\mu_E$ is magnetic-field-like,
\begin{equation}
  h\sim\muB-\mu_E\,,
\end{equation}
and from the fact that singularities in the $h$ plane lie on the
imaginary axis.

The reason that $\muB-\mu_E$ is not $t$-like, but $h$-like,
is the following. In the vicinity of the critical point, since the
O(4) is explicitly broken, the perturbation $\muB-\mu_E$ affects
both the thermal variable $t$ as well as magnetic-field-like variable
$h$ to linear order. Since $\beta\delta>1$, the scaling variable 
$x=th^{-1/(\beta\delta)}$ is small, which means the perturbation
$\muB-\mu_E$ takes the system into the region where the 
variable $h$ dominates the scaling.%
\footnote{An equivalent way of saying this is by comparing the
scaling dimensions of the variables coupled to thermal and magnetic
relevant operators: 
$y_t=1/\nu$ and $y_h=\beta\delta/\nu$. Since 
both operators couple linearly to the variable $\muB-\mu_E$,
and $y_h>y_t$, the magnetic field operator dominates the response to
$\muB-\mu_E$ perturbation near the critical point. In contrast,
at $m=0$, the coupling to magnetic operator is forbidden by symmetry.}

\section{Random matrix model}

In this section we illustrate the universal properties discussed above
by determining the complex plane singularities of a random matrix
model of the QCD partition function at finite $T$ and $\muB$ 
which was introduced in Ref.~\cite{Halasz-pdqcd} and applied
to the study of the QCD phase diagram and the (tri)critical point. At $\mu=0$
this model is equivalent to the finite-$T$ model of 
Ref.~\cite{Jackson-Verbaarschot} and at $T=0$ -- to the finite-$\mu$ model of 
Ref.~\cite{Stephanov-rm}.
The parameters $T$, $\muB$ and $m$ used in this section
are dimensionless, and correspond to
measuring $T$ in units of $T_c=160$ MeV, $\mu$ in units of 2.27 GeV,
and $m$ in units of 100 MeV. For more details, see Ref.~\cite{Halasz-pdqcd}.

The model can be solved in the thermodynamic limit, which corresponds
to the infinite size of the random matrix, $N\to\infty$,
by using the replica trick. This gives, for real $\mu$, $T$ and~$m$,
\begin{equation}\label{zrm}
   \log Z_{RM} =  -\min_{\phi}\Omega(\phi) 
\end{equation}
where
\begin{equation}
\Omega(\phi)=
\phi^2
- \frac12\ln\left\{
[(\phi+m)^2- (\mu+iT)^2]\cdot
[(\phi+m)^2- (\mu-iT)^2]
\right\}.
\end{equation}
Analytically continuing into the complex $\mu$-plane, one finds
the branching points of the partition function by solving
a system of two algebraic equations
\begin{equation}\label{branch}
  \frac{d\Omega}{d\phi}=0;\qquad
  \frac{d^2\Omega}{(d\phi)^2}=0
\qquad \mbox{(branching points)}
\end{equation}
for two unknowns: $\phi$ and $\mu$. The second equation states that
two of the solutions determined by the first equation are coalescing into
one at this value of $\mu$. The Stokes boundaries can be determined
by solving the condition $\Re\Omega(\phi_1)=\Re\Omega(\phi_2)$
where $\phi_1$ and $\phi_2$ are the two solutions of the first
equation in \eqref{branch} which coalesce at the branching point:
\begin{equation}\label{rm-stokes}
    \frac{d\Omega}{d\phi}\Big|_{\phi=\phi_1,\phi_2}=0;\qquad
  \Re\Omega(\phi_1)=\Re\Omega(\phi_2)\,.
\qquad \mbox{(Stokes boundaries)}
\end{equation}

\begin{figure}
\framebox{\epsfig{file=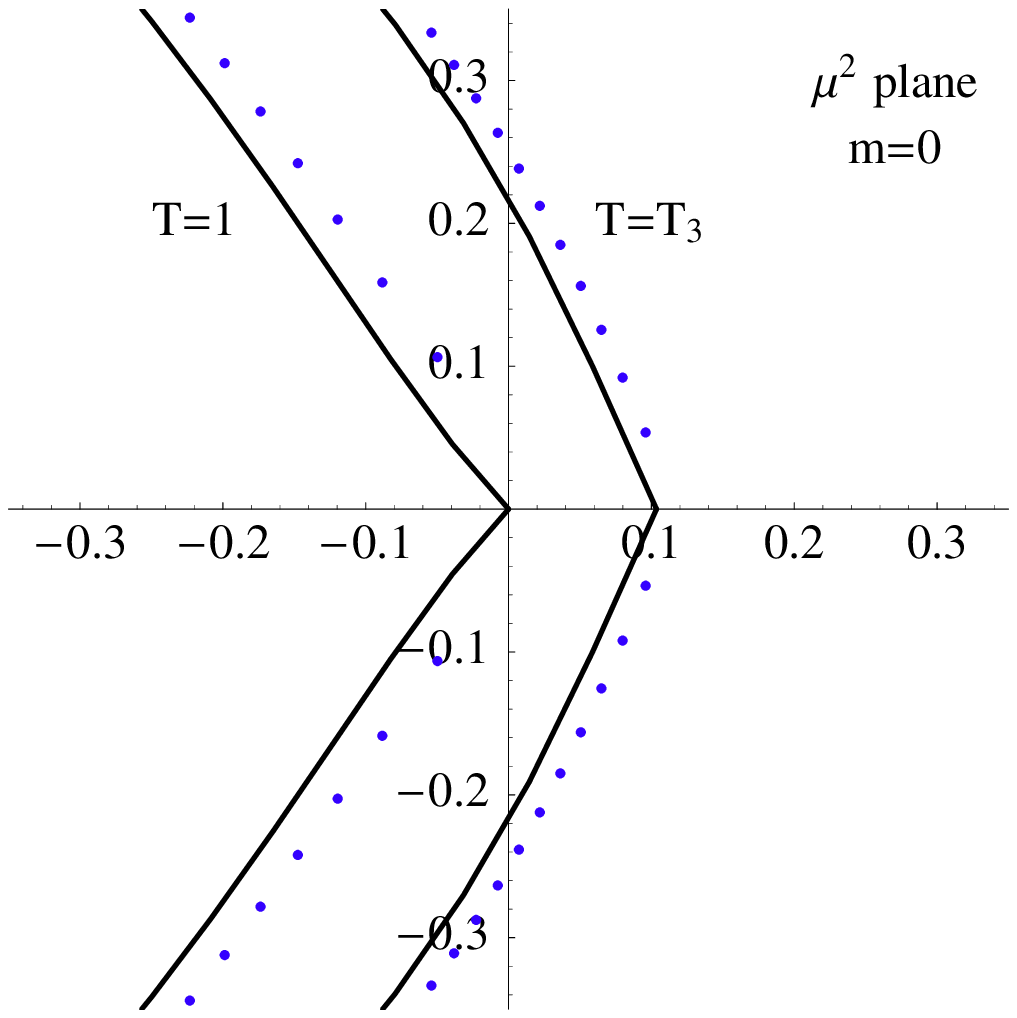,width=0.35\textwidth}}
\hspace{\stretch{1}}
\framebox{\epsfig{file=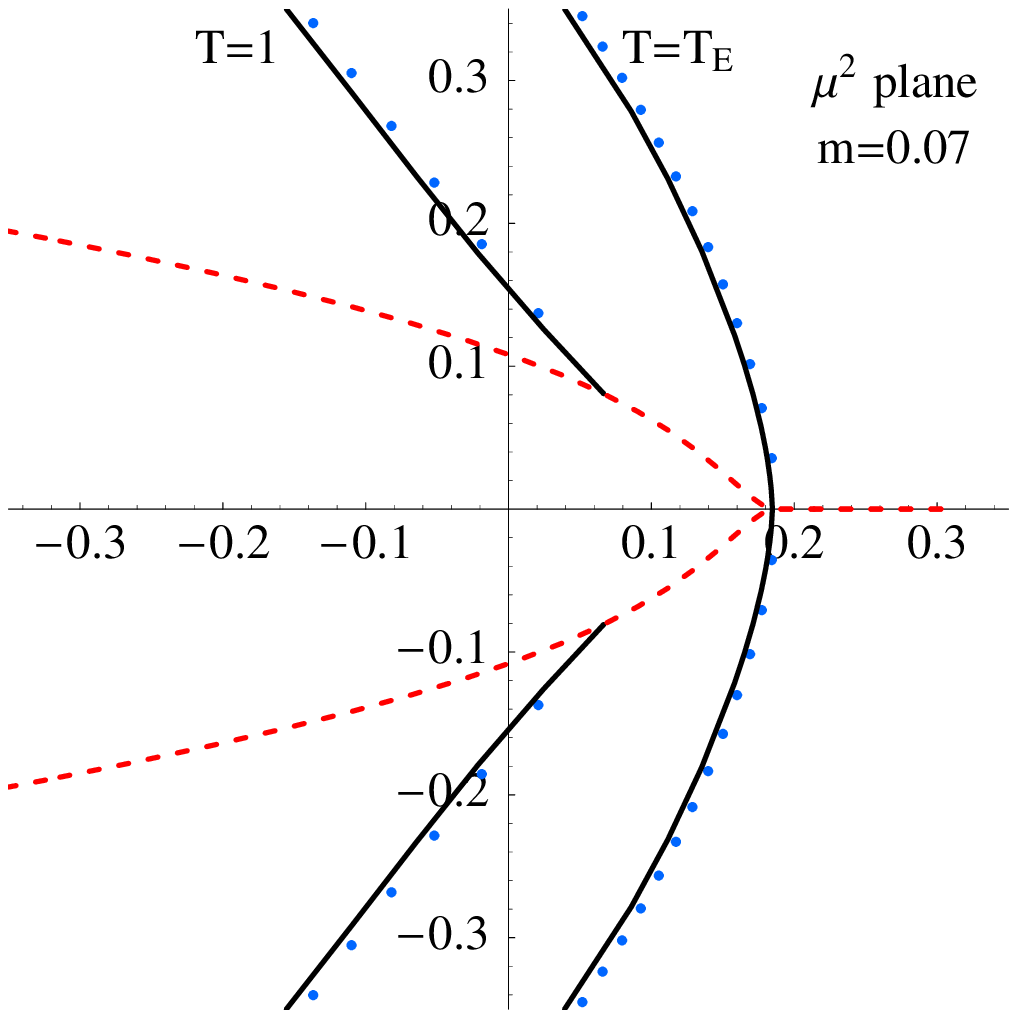,width=0.35\textwidth}}
  \caption[]{Stokes boundaries and zeros of the $N=120$ random matrix
partition function at representative values of $T$ at zero and nonzero
  quark mass. The trajectory of each branching point as a function of $T$
is indicated by a dashed line.}
\label{fig:rm}
\end{figure}

At {\em finite\/} $N$ the partition
function can be written explicitly as a polynomial (apart from an irrelevant
constant factor)
\begin{equation}\label{zrmn}
  \begin{split}
      Z^{(N)}_{RM} = \sum_{k_1,k_2=0}^{N/2}
\binom{N}{k_1}\binom{N}{k_2}
(N-k_1-k_2)!\,
{}_1F_1(k_1+k_2-N;1;-m^2N)
\\\times
\left[-(\mu+iT)^2N\right]^{k_1}\left[-(\mu-iT)^2N\right]^{k_2}
  \end{split}
\end{equation}
using the procedure similar to \cite{Halasz-zeros}, where ${}_1F_1(a;b;c)$
is the Kummer confluent hypergeometric function (in \eqref{zrmn} 
it is a polynomial in
$m^2$). The zeros are found numerically for $N=120$ and plotted in Figure
 \ref{fig:rm} together with the Stokes boundaries given by \eqref{rm-stokes}.

Near the point $T=T_c$, $\muB=0$, the thermal scaling variable
$t$ is proportional to $(T-T_c)+C\muB^2$, where $C$ is the constant
giving the slope the second order transition curve (see
eq.~\eqref{tmu}). Therefore
it is convenient to plot the zeros in the complex $\muB^2$ plane.

Only the vicinity of the origin $\mu^2=0$ is shown in Figure
\ref{fig:rm}. The universal properties
described in the previous section are manifest. It is clear from 
the form of the solution \eqref{zrm} that the critical exponents
near the second order line have their mean field values,
and correspondingly, the angles are $\varphi=45^\circ$ and $\psi=60^\circ$.
At the tricritical point, the exponents are given by their
mean field values also in QCD (albeit with logarithmic corrections):
$\varphi=60^\circ$, $\psi=72^\circ$.
One can also see that the density of the zeros decreases near the
branching point, as dictated by \eqref{rho}.

\begin{figure}
  \epsfig{file=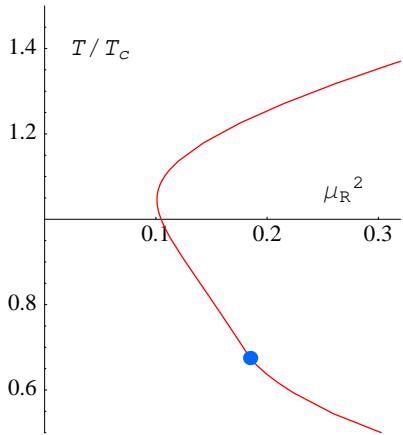,width=0.32\textwidth}
  \caption[]{The convergence radius $\mu_R^2$ as a function of
   $T$ in the random matrix model at $m=0.07$ (7 MeV). 
The value of $\mu_R^2$ is the
distance of the singularity on the dashed line on
  Fig.~\ref{fig:rm} from the origin. 
The critical point at $T=T_E$
where the singularities pinch
the real axis is shown.}
\label{fig:murt}
\end{figure}

\section{Convergence radius}

Relevant for the search of the QCD critical point is the question of
the convergence radius $\mu^2_R(T)$ 
of the Taylor expansion around $\mu=0$ of the QCD
thermodynamic potential as a function of $T$.
For sufficiently small quark masses, and sufficiently near $T_c$,
the position of the nearest singularity, limiting this radius,
is determined by the universal arguments given above. For
illustration,  this radius in the random matrix
model is plotted on Figure~\ref{fig:murt}. 

All generic and universal features are manifest in Fig.~\ref{fig:murt}.
As $T$ decreases, and the branching point singularity slides along 
the dashed line on Fig.~\ref{fig:rm} from left to right,
the radius $\mu_R$ contracts, and then, below the crossover
temperature, begins to expand again.
From the universality arguments of Section \ref{sec:qcd-universal}
(see Fig. \ref{fig:qcdzeros} (right)) we conclude that near the chiral
($m\to0$) limit the minimum value of the radius scales with $m$ as
\begin{equation}
\min_T\, \mu_R^2(T) \sim m^{1/(\beta\delta)} \sim m^{0.54}
\end{equation}
and is achieved at a temperature $T$ which scales as 
$T-T_c\sim m^{1/(\beta\delta)}\sim m^{0.54}$.

Further away from the minimum, at the critical point, $T=T_E$,
the singularity and its conjugate pinch the real axis. At
this point one observes a non-analyticity in $\mu_R^2(T)$:
a contribution of order 
$(T_E-T)^{\beta\delta}$  turns on
below $T=T_E$. 
This is due to the kink in the dashed line
on Fig.~\ref{fig:rm} at $T_E$. In QCD $\beta\delta\approx 1.56$,
given by the exponents of the 3d Ising model, while in the random matrix
model $\beta\delta$ has the mean field value $3/2$.
More explicitly, the trajectory of the
branching point $\mu_*$ near $\mu_E$ is given by:
\begin{equation}\label{mu*}
  \mu_*(T)=\mu_E+c_1(T_E-T)+i\,c_2(T-T_E)^{\beta\delta}
+{\cal O}\left((T-T_E)^2\right),
\end{equation}
with some nonuniversal positive coefficients $c_{1,2}$. To derive this
equation one observes that both $t$ and $h$ scaling variables
are linear combinations of $(T-T_E)$ and $(\mu-\mu_E)$ and
uses equation \eqref{psi} for the branching point. The fact that
the third term on the r.h.s. in \eqref{mu*} is purely imaginary for $T>T_E$
is related to the fact that the branching point $h_*$ in the $h$
plane is on the imaginary axis for $t>0$ as discussed in 
Section~\ref{sec:hne0}. Therefore
\begin{equation}\label{murte}
  \mu_R^2(T)=|\mu_*(T)|^2=\mu_E^2+\tilde{c}_1(T_E-T)
+ \tilde c_2\,\theta(T_E-T)(T_E-T)^{\beta\delta} 
+ {\cal O}\left((T-T_E)^2\right).
\end{equation}

Below $T_E$,
the singularity continues to move away from the origin, and the radius
of convergence continues to increase. The radius is now determined by the
spinodal point of the first order phase transition (but see discussion
in Appendix).
This singularity resides on the continuation
of the physical Riemann sheet under the cut.%
\footnote{On a more subtle level, one has to note that
the fact that the singularity \eqref{mu*} 
remains on the real axis is an artifact
of the mean field critical behavior in the random matrix model 
($\beta\delta=3/2$).
In QCD, the singularity moves off the real axis by an amount which
scales as $(T_E-T)^{\beta\delta}$.
} 

One must point out that the random matrix model of
Ref.~\cite{Halasz-pdqcd} does not capture a
known feature of the QCD partition function -- the periodicity, or invariance 
under the shift $\muB\to\muB+2\pi iT$, which is due to the quantization
of the baryon charge.\footnote{A random matrix model which does capture this
feature is studied in \cite{Halasz-matsubara}.}
As shown by Roberge and Weiss \cite{Roberge}
this periodicity is related to the appearance of a
Stokes boundary given by $\Im \muB=\pi T$ for sufficiently
high temperatures. For $T$ of order 160 MeV, this Stokes boundary could
interfere with convergence of the series only if the singularity
we discuss moves further than $|\muB| \approx 500$~MeV.

\section{Summary and discussion}

We have described the location as well as temperature and quark mass
 dependence of the singularities
of the QCD partition function in the complex $\muB$ plane.
In the vicinity of the chiral phase transition at $\mq=0$ the universality
and scaling arguments predict that in the infinite volume
the singularities are two complex conjugate branch cuts
originating at a branching point on 
the real $\mu$ axis. The cuts are oriented at an angle to the
negative $\mu$ axis given by \eqref{phi}. At nonzero $m$ the branching
points (and the cuts) are shifted in the direction given by angle 
$\psi\approx 48^\circ$, by a distance of order $m^{0.54}$ 
(see Fig.~\ref{fig:qcdzeros}).

A related consequence of the universal behavior of the complex $T$
singularities, and the fact that $\psi<90^\circ$, is the prediction
that the
crossover point at $m\ne0$, defined as the projection of the closest
singularity onto the real axis, is {\em above} the second order O(4)
line, as sketched in Fig. \ref{fig:pd-sketch}.

The singularities we describe determine the convergence of the 
Taylor expansion around the point $\mu=0$.
As a result, the radius of convergence $\muR^2$ at $T=T_c$
is limited by a singularity whose distance from the origin
scales as $\muR^2\sim m^{0.54}$, vanishing in the chiral limit.
At $T=T_E$ the convergence radius $\mu_R(T)$ shows nonanalyticity
described by Eq.~\eqref{murte}.

The random matrix model of Ref.~\cite{Halasz-pdqcd} illustrates
these universal predictions: Figures \ref{fig:rm}, \ref{fig:murt}.

The knowledge of the complex plane singularities might be used to
improve the Taylor expansion methods, for example, by constructing
Pad\'e or similar extrapolations \cite{Lombardo}, accommodating the
correct universal singular behavior. It can also be used to crosscheck the
results of lattice Monte-Carlo simulations, by comparing the expected
universal behavior of the partition function 
zeros to the output of a lattice calculation.

\begin{acknowledgments}
  This work was supported by the DOE grant No.\ DE-FG0201ER41195, and
  by the Alfred~P.~Sloan Foundation.
\end{acknowledgments}

\appendix*

\section{Taylor expansion near a Stokes boundary}

The following question has the practical importance for lattice Taylor
expansion methods \cite{Allton,Gavai,Lombardo}: 
Does a Stokes boundary limit the convergence radius of
a Taylor expansion and how? A view often expressed in the
literature is that the thermodynamic functions can be analytically
continued past the Stokes boundary (or a first order phase
transition) -- the true singularities being located at the branching
points, where the Stokes boundaries end. Although this is correct
in the strictly infinite volume $V=\infty$, 
this certainly is not correct for any finite
volume, no matter how large. This must be so since the
singularities of the free energy of the type $\log(z-z_k)$ 
are located along the Stokes
boundaries. How then do these singularities appear {\em not\/}
 to limit the radius of
convergence in the $V\to\infty$ limit? The answer is that they do,
but in order to see the divergence in the series, one needs to continue
the expansion beyond some large order $n_*$ which slides to infinity
as $V\to\infty$. We determine $n_*$ in this Appendix.

Consider a simplified problem, where the expansion point $z_0$ 
is located sufficiently close to the Stokes boundary so that the
curvature of the boundary can be neglected, and the zeros
can be considered equally spaced and extending to infinity. 
It is also convenient to
transform the variable $z$ to conformally 
map the Stokes boundary to the imaginary
axis as shown in Figure~\ref{fig:taylor-z0}.

 The contribution to
the thermodynamic potential from the Stokes boundary at finite volume
is the given by
\footnote{
A more common example $\Omega=-\log2\cosh(zV)$ can be analyzed
similarly, with a little additional complication, unnecessary for this
discussion.
}
\begin{equation}
  \Omega_{\rm sing} = -\log\left\{2Vz\prod_{k=1}^{\infty}\left[1-\left(\frac{zV}{\pi
  k}\right)^2\right]\right\} = -\log2\sinh(zV) 
= - zV - \ln\left(1-e^{-2zV}\right)\,.
\end{equation}
We have also rescaled the variable $z$ so that the spacing between zeros
is $\pi/V$, to emphasize the fact the the density of zeros grows
linearly with $V$.
 In the thermodynamic limit
\begin{equation}
  \frac {\Omega_{\rm sing}} V \,\stackrel{V\to\infty}{\xrightarrow{\phantom{V\to\infty}}}\,
\left\{
\begin{array}{ll}
- z, &\quad \Re z>0\,;\\
\phantom{-} z+i\pi, &\quad \Re z<0\,.
\end{array}
\right.
\end{equation}
On the Stokes boundary, $\Re z=0$, $\Re\Omega$ is continuous but the
normal derivative of $\Re\Omega$
is not, in accordance with the electrostatic analogy \eqref{E}. 
However, the function
$-z$ has no singularity on the Stokes boundary and can be
analytically continued through it.
Let us choose $z_0$ to be on the positive real axis 
and Taylor expand around such a point (see
Fig.~\ref{fig:taylor-z0}): 
\begin{figure}
  \epsfig{file=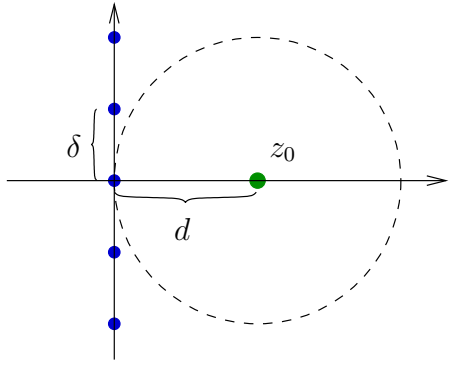}
  \caption[]{The convergence disk of the Taylor series around point
    $z_0$ near a line of complex plane singularities.}
\label{fig:taylor-z0}
\end{figure}
\begin{equation}
  \Omega_{\rm sing}(z)-\Omega_{\rm sing}(z_0)=-(z-z_0)V + \sum_{n=1}^\infty a_n
  \left(\frac{z_0-z}{z_0}\right)^n
\end{equation}
where
\begin{equation}
a_n = \frac1{n!}\,(2z_0V)^n\,\sum_{p=1}^\infty p^{n-1}\,e^{-2z_0Vp}.
\end{equation}
For large $V$ and $n$ the coefficients approach%
\footnote{In the interval $n_*\ll n\ll n_*^2/(2\pi)$ the coefficients
oscillate with maxima reaching $n_*/\sqrt{2\pi n^3}$
 before settling on $a_n=1/n$.}
\begin{equation}
a_n\, \stackrel{n\to\infty,\, V\to\infty}
{\xrightarrow{\phantom{n\to\infty,\,V\to\infty}}}
\left\{  \begin{array}{ll}
e^{-n_*}\,(n_*)^n/n!\,
,\qquad & n\ll n_* \equiv 2z_0V\,;
\\
1/n\,
,\quad& n\gg n_{*}\, .
  \end{array}\right.
\end{equation}
We see that the late terms $n>n_*\sim V$ will cause the
divergence of the Taylor series outside the circle of radius $|z_0|$.
The value of $n_*$ can be written in terms of two quantities:
the distance $d$ of the expansion point from the Stokes boundary and
the spacing $\delta$ between the zeros at the given volume (shown in
Fig.~\ref{fig:taylor-z0}). Recalling
that in terms of the rescaled variable~$z$ the spacing is $\delta=\pi/V$ and
the distance is $d=|z_0|$, we conclude:
\begin{equation}
  n_*= \frac{2\pi d}{\delta}\,.
\end{equation}

\end{document}